\documentclass[12pt,preprint]{aastex}
\usepackage{emulateapj5}

\newcommand{\Msol}{M$_{\odot}$}
\newcommand{\Mbol}{M$_{bol}$}
\newcommand{\Mjup}{M$_{\mathrm{JUP}}$}

\newcommand{\Msini}{M\,$\sin i$}

\newcommand{\ms}{m\,s$^{-1}$}

\newcommand{\rhk}{R$^{\prime}_{HK}$}
\newcommand{\au}{a.u.}
\received{November 2001}
\accepted{}
\journalid{}{}
\slugcomment{Submitted to  {\em The Astrophysical Journal}.}

\lefthead{Tinney et al.}
\righthead{Planets from the Anglo-Australian Planet Search}

\begin{document}
\title{Two extra-solar planets from the Anglo-Australian Planet Search\altaffilmark{1} }

\author{C.G. Tinney\altaffilmark{2}, R. Paul Butler\altaffilmark{3}, 
        Geoffrey W. Marcy\altaffilmark{4,5}, Hugh R.A. Jones\altaffilmark{6}, 
        Alan J. Penny\altaffilmark{7}, Chris McCarthy\altaffilmark{3}, 
        Brad D. Carter\altaffilmark{8}}

\altaffiltext{1}{Based on observations obtained at the
    Anglo--Australian Telescope, Siding Spring, Australia.}
\altaffiltext{2}{Anglo-Australian Observatory, PO Box 296, Epping. 1710. 
Australia. {\tt cgt@aaoepp.aao.gov.au}}
\altaffiltext{3}{Carnegie Institution of Washington,Department of Terrestrial Magnetism,
       5241 Broad Branch Rd NW, Washington, DC 20015-1305}
\altaffiltext{4}{Department of Astronomy, University of California, Berkeley, CA, 94720}
\altaffiltext{5}{Department of Physics and Astronomy, San Francisco State University, San Francisco, CA 94132.}
\altaffiltext{6}{Astrophysics Research Institute, Liverpool John Moores University, 
       Twelve Quays House, Egerton Wharf, Birkenhead CH41 1LD, UK}
\altaffiltext{7}{Rutherford Appleton Laboratory, Chilton, Didcot, Oxon OX11 0QX, U.K.}
\altaffiltext{8}{Faculty of Sciences, University of Southern Queensland, Toowoomba, 4350. Australia.}

\begin{abstract}
We report the detection of two new extra-solar planets
from the Anglo-Australian Planet Search around the stars HD\,142
and HD\,23079. The planet orbiting HD\,142 has an orbital 
period of just under one year,
while that orbiting HD\,23079 has a period of just under two years.
HD\,142 falls into the class of ``eccentric''
gas giants. HD\,23079 lies in the recently uncovered class of
``$\epsilon$\,Ret-like'' planets -- extra-solar gas giant planets with near-circular orbits outside 0.1\,\au\
The recent discovery of several more members of this
class provides
new impetus for the extension of existing planet searches to longer
periods, in the search for Jupiter-like planets in Jupiter-like orbits.

\end{abstract}


\section{The Anglo-Australian Planet Search}

The Anglo-Australian Planet Search (AAPS) is a long-term planet
detection program which aims to perform extra-solar
planet detection and measurement at the highest possible precision.
Together with programmes using similar techniques on the
Lick 3\,m and Keck I 10\,m telescopes \citep{fischer01,vmba00}, it provides all-sky planet search coverage for inactive F,G,K and M dwarfs
down to a magnitude limit of V=7.5. Initial results from
this programme demonstrate that AAPS acheives long-term,
systematic velocity precisions of 3\,\ms\ or better 
\citep{tbjpvah01,btmjpa01}.

AAPS is being carried out on the 3.92\,m Anglo-Australian
Telescope (AAT), using the University College of London Echelle Spectrograph (UCLES)
and an I$_2$ absorption cell.
UCLES is operated in its 31\,lines\,mm$^{-1}$ mode. Prior to 2001 September,
it was used with  a MIT/LL 2048$\times$4096 15$\mu$m pixel CCD, and since then has
been used with an EEV 2048$\times$4096 13.5$\mu$m pixel CCD. 
Our target sample includes 178 FGK stars with $\delta < -20$\arcdeg\ and V$<$7.5,
and a further 23 M and metal enriched stars with V$<$11.5. 
Where age/activity information is available 
from \rhk\ indices (see e.g., \citet{hsdb96,tmjbcmp02}) we require
 target stars to have \rhk $<$ $-4.5$ (corresponding to ages
greater than 3\,Gyr). The observing and data processing 
procedure follows that described in \citet{bmwmd96} and  
\citet{btmjpa01}.

\section{Characteristics of HD\,142 \& HD\,23079}

\begin{figure}
\epsscale{0.5}
\plotone{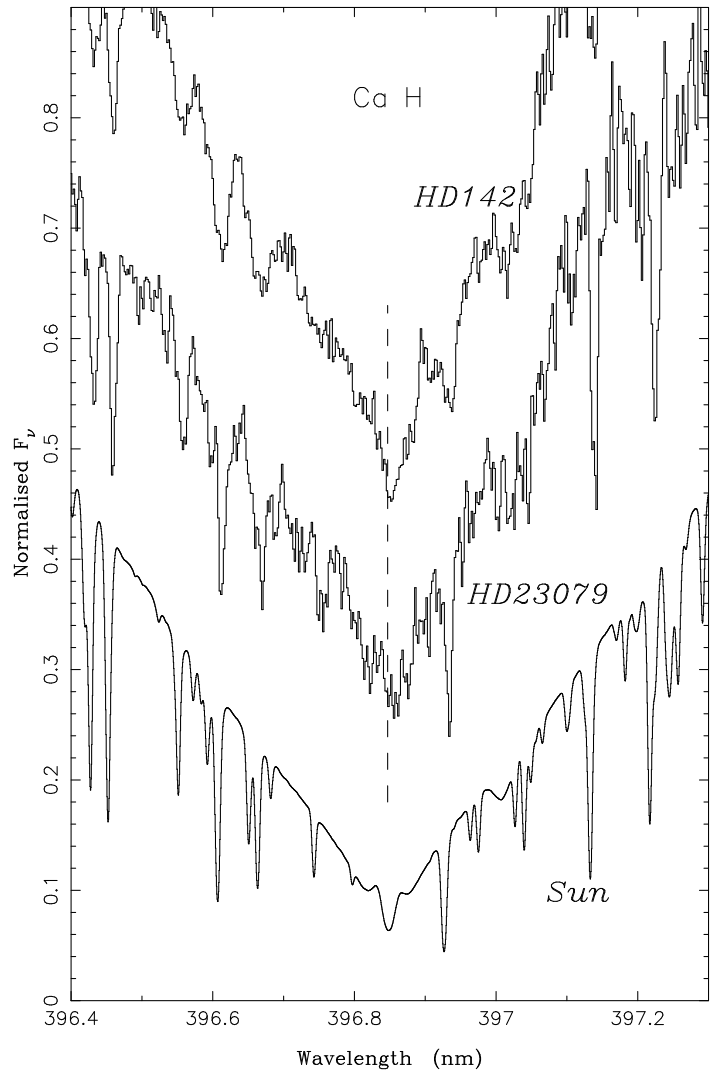}
\caption{Comparison of the \ion{Ca}{2} H line core in the Sun (lower line),
  HD\,23079 (middle line) and HD\,142 (upper line). Solar spectrum is from
  \citet{kurucz84} and other spectra are from \citet{tmjbcmp02}.  }
\label{Ca_line}
\end{figure}

HD\,142 (HR\,6, HIP\,522, GJ\,4.2A, LHS\,1020) is a chromospherically inactive 
({\rhk}=$-4.92$) G1IV star \citep{mssII,tmjbcmp02}.
Its Hipparcos parallax of 39.0$\pm$0.6\,mas implies absolute magnitudes
of M$_V$=3.66$\pm$0.03 \citep{esa97}
and \Mbol=5.55$\pm$0.05 \citep{alonso95} .
%
%
%
%
The fundamental parameters of HD\,142 have been examined via
spectroscopy \citep{favata97} and Str\"omgren $ubvy$ photometry
(see the compilation of \citet{eggen97}). Spectroscopy
derives [Fe/H]=$+0.04{\pm}0.15$ and T$_{\mathrm eff}$=6025\,K, 
while the photometry suggests [Fe/H]=$-0.04$, which is in
agreement with the spectroscopy to within uncertainties.
Based on interpolation
between the evolutionary tracks of \citet{fpb98,fpb97}, the 
mass of HD\,142 is estimated to be 1.15$\pm$0.1\Msol.
Figure \ref{Ca_line} shows the \ion{Ca}{2} H line region for HD\,142, together
with the quiet Sun and HD\,23079. The absence of significant emission
confirms that this star is chromospherically inactive. 

\placefigure{Ca_line}

%
%

HD\,23079 (HIP\,17096, LTT\,1739) is an  inactive dwarf with a {\rhk}=$-4.96$
\citep{tmjbcmp02,hsdb96}. \citet{mssI} classify it as F8/G0V 
(i.e. intermediate between F8 and G0).
Its Hipparcos parallax is 28.9$\pm$0.6\,mas,
giving it M$_V$=4.42$\pm$0.05 and \Mbol=4.25$\pm$0.05 \citep{esa97,lang}. 
%
%
%
%
No metallicity information is available for this star,
so mass estimation will be less precise.
At [Fe/H]=+0.25, 0.0  and $-0.25$,
the models of \citet{fpb98,fpb97} would indicate M=1.25, 1.10 and 1.0\,\Msol\ (respectively).
For the most likely metallicity range of this F/G-dwarf, its mass lies in
the range 1.0-1.25\,\Msol, and we therefore adopt  M=1.10$\pm$0.15\,\Msol.
Both HD\,142 and HD\,23079 were seen to be photometrically stable over the life of the Hipparcos mission
at a 95\% confidence level of $<$0.018 magnitudes \citep{esa97}.
%
%

\section{Radial Velocity Observations and Orbital Solutions}
\label{velocities}

\begin{figure}
\epsscale{0.5}
\plotone{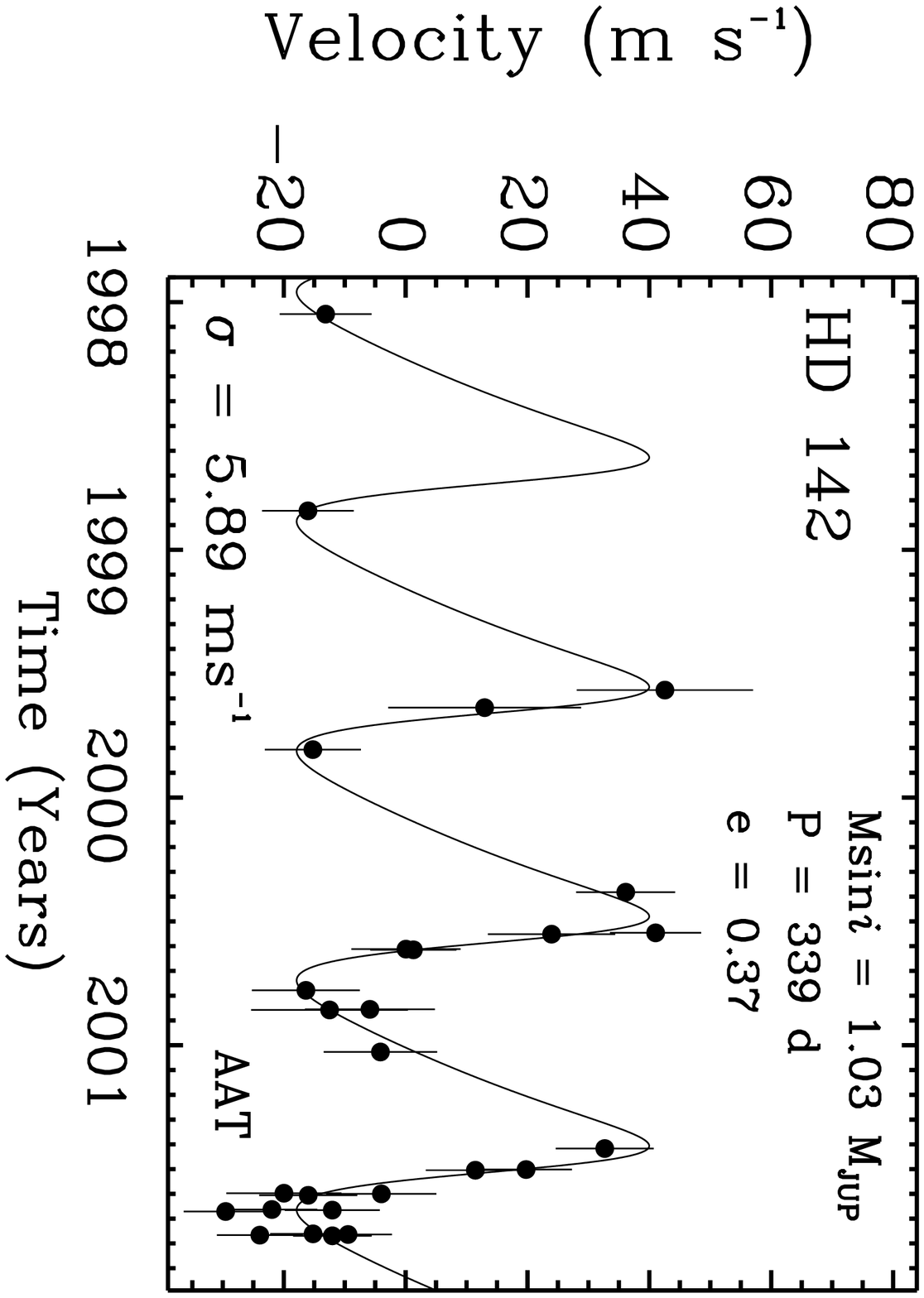}
\caption{AAT Doppler velocities for HD\,142 from 1998 January to
2001 October.   The solid line is a best
fit Keplerian with the parameters shown
in Table \ref{orbits}.The rms of the velocities about the fit is 5.89\,\ms. 
Assuming 1.15\,\Msol\ for the primary,
the minimum (\Msini) mass of the companion is 1.03$\pm$0.19\,\Mjup, and
the semimajor axis is 1.0$\pm$0.1\,au.}
\label{hd142_rv_curve}
\end{figure}

Twenty-seven observations of HD\,142 are listed in Table \ref{vel142}. The 
column labelled ``Uncertainty'' is the velocity uncertainty produced by our
least-squares fitting. This fit simultaneously determines the Doppler shift and
the spectrograph point-spread function (PSF) for each observation
made though the iodine cell, given an iodine absorption spectrum and
an ``iodine free'' template spectrum of the object \citep{bmwmd96}. 
The uncertainty is derived from the ensemble of velocities from
each of four hundred useful spectral regions (each 2\,\AA\ long) in every exposure.
This uncertainty includes the effects of photon-counting uncertainties,
residual errors in the spectrograph PSF model, and variation in the underlying 
spectrum between the template and ``iodine'' epochs. All velocities are measured relative to the zero-point defined by the template observation. 
Only observations
where the uncertainty is less than twice the median uncertainty are
listed.
These data are shown in Figure \ref{hd142_rv_curve}. 
The figure shows the best-fit Keplerian model for the
data, with the resultant orbital parameters listed in Table \ref{orbits}. 

The residuals about the fit are slightly higher than
the 3-4\,\ms\ average level of ``jitter'' expected in a G1 star with HD\,142's level
of activity \citep{sbm98}, but is within the typical range seen in
even inactive stars\footnote{``Jitter'' here is used to refer to the scatter in the observed
velocity about a mean value in systems observed over the long-term to 
have no Keplerian Doppler shifts, or about a fitted Keplerian in systems
known to have a planetary mass companion. It is thought to be due to the combined
effects of surface inhomogeneties, stellar activity and stellar rotation.}.

\placetable{vel142}  

\placefigure{hd142_rv_curve} 

\placetable{orbits}  

\begin{figure}
\epsscale{0.5}
\plotone{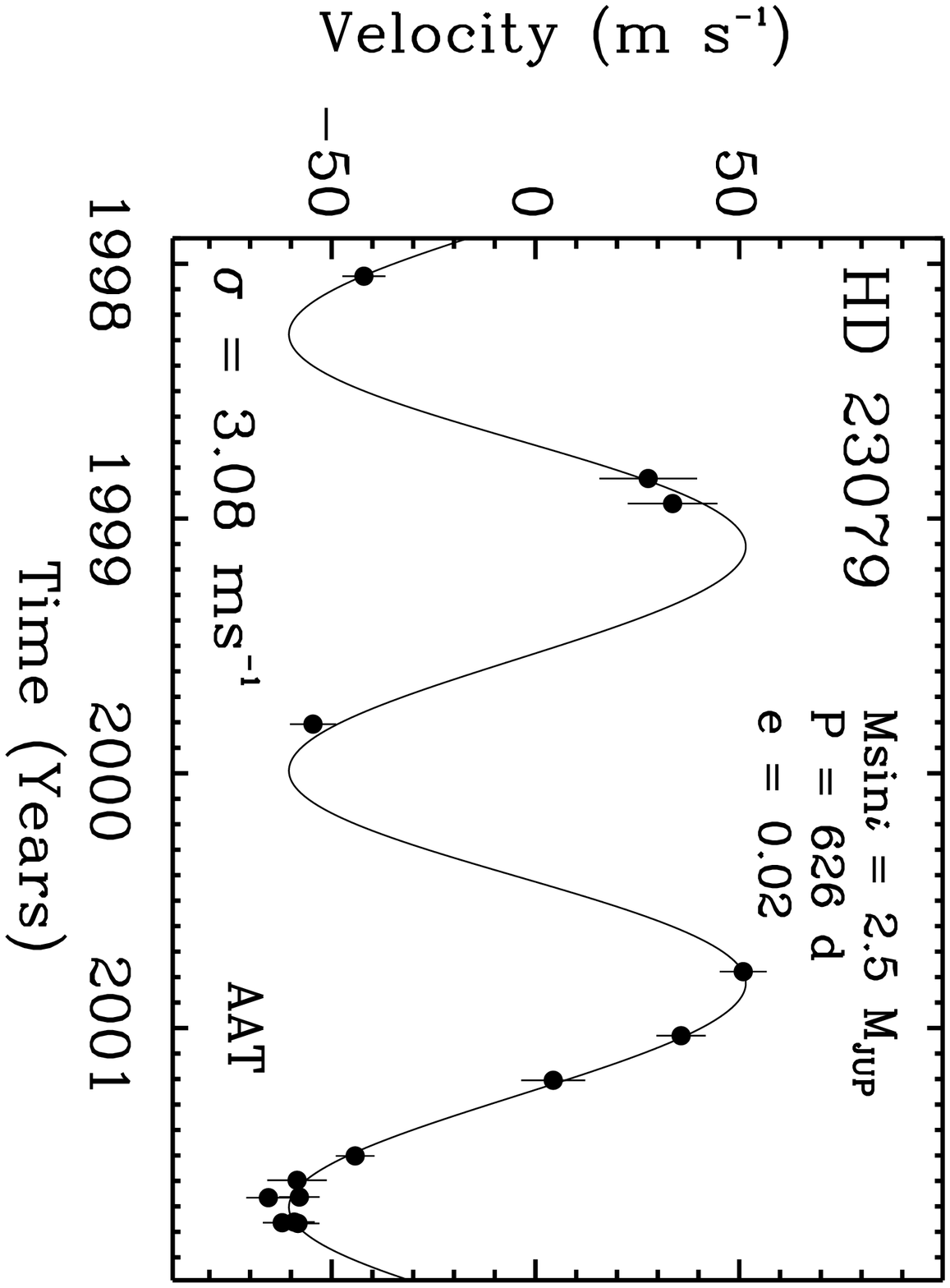}
\caption{AAT Doppler velocities for HD\,23079 from 1998 January to
2001 October. The solid line is a best fit Keplerian with the parameters shown
in Table \ref{orbits}.
The rms of the velocities about the fit is 3.08\,\ms. Assuming a 1.1\,\Msol\ for the primary,
the minimum (\Msini) mass of the companion is 2.5$\pm$0.3\,\Mjup, and
the semimajor axis is 1.5$\pm$0.2\,au.}
\label{hd23079_rv_curve}
\end{figure}

The thirteen observations of HD\,23079 are listed in Table \ref{vel23079}, 
and they are shown in Figure \ref{hd23079_rv_curve}
along with a Keplerian fit to the data with
the orbital parameters listed in Table \ref{orbits}. The rms scatter about this fit 
of only 3.08\,\ms\ demonstrates the extra-ordinary control over long term
systematics which the iodine cell technique can deliver for stars with
suitable intrinsic velocity stability. It also demonstrates the
suitability of the UCLES spectograph at the AAT for radial velocities
at the highest precisions -- even for a V=7.1 star near our
V=7.5 current survey limit.

\placetable{vel23079}  

\placefigure{hd23079_rv_curve} 

\section{Discussion}

The resultant minimum companion mass for HD\,142 is \Msini\,=\,1.03$\pm$0.19\,\Mjup,  
with an orbital semi-major axis $a$\,=\,1.0$\pm$0.1\,au\ 
at an eccentricity of 
$e$\,=\,0.37$\pm$0.1 -- a roughly Jupiter-mass giant planet in an Earth-like, but eccentric,
orbit. An interesting feature of HD\,142 is that its metallicity is only
marginally enriched over solar. This planet, at least, has not formed in
a metal-enriched system, as has been suggested for many
of the extra-solar planets (e.g. see \citet{gonzalez01} and references therein).

\begin{figure}
\epsscale{0.9}
\plotone{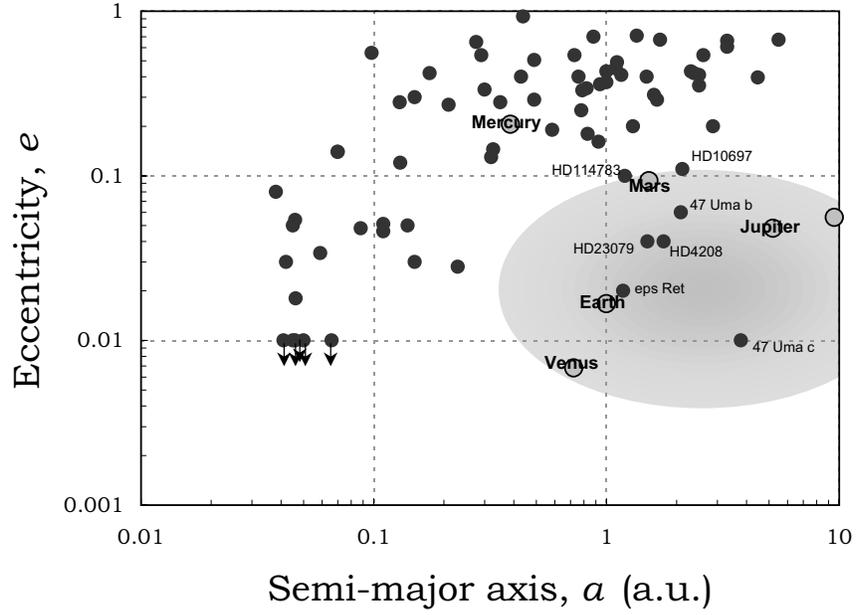}
\caption{Log($e$) versus Log($a$) for extra-solar planets reported
as of 2001 September, plus the new planets reported in this paper and by
\citep{vbmfpal02} {\em (solid circles)}, together with the
inner planets of the Solar System {\em (shaded circles)}. Planets with measured eccentricities $e < 0.01$ are shown 
as upper limits at $e = 0.01$. The region of the Log($e$)-Log($a$)
occupied by the planets of the Solar System, and its similarity to
those of the ``$\epsilon$\,Ret-like'' planets is highlighted.}
\label{eversusa}
\end{figure}

The minimum companion mass and orbital parameters
derived for HD\,23079 (\Msini =2.5$\pm$1.1\,\Mjup,  $a$=1.5$\pm$0.2\,au, 
$e$=0.04$\pm$0.18) indicate the presence of a planet with significantly
larger mass than Jupiter, in a Mars-like orbit with eccentricity
consistent with zero. Together with the extra-solar planetary
companions to $\epsilon$\,Ret \citep{btmjpa01}, HD\,4208 \citep{vbmfpal02}, 
47\,UMa\,b\,\&\,c \citep{fischer02}, 
and possibly the companions to HD\,114783 and
HD\,10697 \citep{vbmfpal02,vmba00}, HD\,23079 forms a new class
of extra-solar planets, which we name after the prototype
object $\epsilon$\,Ret. The region of the Log($e$) versus Log($a_{maj}$)
diagram these ``$\epsilon$\,Ret-like'' planets occupy is highlighted in Fig. \ref{eversusa}.
It is worth remembering that prior to about 12 months ago the highlighted
region of this plot was empty -- though many extra-solar
planets had been discovered, none
shared orbital properties with the planets of our own Solar System.
The ``$\epsilon$\,Ret-like'' planets, therefore, join with
the ``51\,Peg-like'' and eccentric giant
planets in filling out the bestiary of extra-solar planets. Prior to their
detection, it was unclear whether giant planets in circular,
or near-circular, orbits outside 0.1\,\au\, would be found {\em at all} outside the Solar System.
Their discovery points the way to the detection of Solar System analogs (in
the form of Jupiter-like-planets in Jupiter-like-orbits) once data sequences
at better than 2-3\,\ms\ span the necessary 10-12 year periods.

\section{Conclusions}

We present results for the detection and characterisation for two new
extra-solar planets with orbital periods of one year or greater
around the
stars HD\,142 and HD\,23079. The planet around HD\,23079
is particularly interesting --  it represents the detection of
a new member of the class of ``$\epsilon$\,Ret-like''
giant planets in near-circular orbits outside 0.1\,\au. The continued
detection by high precision Doppler searches of these gas giants,  in
Solar System-like orbits, gives added impetus that the continuation
of these searches to the 10-12 year periods where analogs of the gas giants
in our own Solar System may become detectable around other stars.

\acknowledgments
The Anglo-Australian Planet Search team would like to gratefully acknowledge the support
of Dr Brian Boyle, Director of the AAO,  and the superb technical support which has been
received throughout the programme from AAT staff - in particular E.Penny, R.Patterson, D.Stafford, F.Freeman, S.Lee, J.Pogson and G.Schafer. We further acknowledge support 
by; the partners of the Anglo-Australian Telescope Agreement (CGT,HRAJ,AJP);
    NASA grant NAG5-8299 \& NSF grant AST95-20443 (GWM);
    NSF grant AST-9988087 (RPB); and Sun Microsystems.
NSO/Kitt Peak FTS data used here were produced by NSF/NOAO.

\begin{deluxetable}{lrr}
\tablenum{1}
\tablecaption{Velocities for HD\,142}
\tablewidth{0pt}
\tablehead{
JD$^a$ & RV$^a$ & Uncertainty \\
(-2450000)   &  (\ms) & (\ms)
}
\startdata
   830.9587  &   -13.1  &  7.5 \\
  1121.0194  &   -16.0  &  7.5 \\
  1385.3105  &    42.6  & 14.5 \\
  1411.2025  &    13.0  & 15.8 \\
  1473.0850  &   -15.2  &  7.9 \\
  1683.3314  &    36.1  &  8.1 \\
  1743.2765  &    41.0  &  7.5 \\
  1745.2642  &    24.0  & 10.4 \\
  1767.2699  &     0.1  &  9.0 \\
  1768.2542  &     1.3  &  7.1 \\
  1828.0607  &   -16.4  &  8.9 \\
  1856.0643  &    -5.9  & 10.7 \\
  1856.9250  &   -12.5  & 12.9 \\
  1918.9407  &    -4.1  &  9.3 \\
  2061.2963  &    32.7  &  8.0 \\
  2092.2683  &    19.8  &  7.5 \\
  2093.2876  &    11.4  &  8.1 \\
  2127.2230  &   -20.0  &  9.4 \\
  2128.1545  &    -4.0  &  9.0 \\
  2130.2433  &   -16.0  &  8.0 \\
  2151.2113  &   -21.9  &  7.4 \\
  2152.0786  &   -12.0  &  7.8 \\
  2154.1541  &   -29.5  &  6.9 \\
  2187.1000  &   -15.2  &  7.0 \\
  2188.0360  &    -9.4  &  7.2 \\
  2189.0199  &   -23.9  &  7.0 \\
  2190.0032  &   -12.0  &  6.4 \\
\enddata
\tablenotetext{a}{Radial Velocities
(RV) are barycentric, but have an arbitrary zero-point determined by
the radial velocity of the template, as described in Section~\ref{velocities}}
\label{vel142}
\end{deluxetable}

\clearpage

\begin{deluxetable}{lcc}
\tablenum{2}
\tablecaption{Orbital Parameters}
\tablewidth{0pt}
\tablehead{
\colhead{Parameter}            & \colhead{HD\,142} 
                                                   & \colhead{HD\,23079}
                                                                      \\
}
\startdata
Orbital period $P$ (d)          &   339$\pm$6     &  626$\pm$24       \\
Velocity amp. $K$ (\ms)         &    29.6$\pm$5     &  56$\pm$5        \\
Eccentricity $e$                &  0.37$\pm$0.1     & 0.02$\pm$0.12     \\
$\omega$ (\arcdeg)              &    71$\pm$36      & 262$\pm$50        \\
$a_1 \sin i$ (km)               &  (0.1280$\pm$0.0066)$\times$10$^6$
                                                    & (0.482$\pm$0.019)$\times$10$^6$\\
Periastron Time (JD-2450000)     & 1752$\pm$22       &  1680$\pm$90     \\
\Msini\ (\Mjup)                 &  1.03$\pm$0.19    &  2.5$\pm$0.3     \\
a (AU)                          &  1.0$\pm$0.1      &  1.5$\pm$0.2      \\
RMS about fit (\ms)             &  5.89              &  3.08              \\
\enddata
\label{orbits}
\end{deluxetable}

\begin{deluxetable}{lrr}
\tablenum{3}
\tablecaption{Velocities for HD\,23079}
\tablewidth{0pt}
\tablehead{
JD$^a$ & RV$^a$ & Uncertainty \\
(-2450000)   &  (\ms) & (\ms)
}
\startdata
   831.0689  &   -42.1  &  5.3 \\
  1121.1268  &    27.7  & 12.0 \\
  1157.0594  &    33.6  & 11.0 \\
  1473.2492  &   -54.6  &  5.6 \\
  1828.1399  &    51.0  &  5.8 \\
  1920.0142  &    35.7  &  6.0 \\
  1983.8858  &     4.4  &  7.8 \\
  2092.3211  &   -44.2  &  4.7 \\
  2127.2797  &   -58.5  &  7.3 \\
  2151.2764  &   -57.9  &  5.0 \\
  2152.2093  &   -65.5  &  5.4 \\
  2187.1679  &   -59.2  &  5.0 \\
  2188.1270  &   -62.1  &  4.7 \\
  2189.1418  &   -58.3  &  5.2 \\
\enddata
\tablenotetext{a}{As for Table \ref{vel142}}
\label{vel23079}
\end{deluxetable}

\clearpage

\end{document}